\def\year{2022}\relax
\documentclass[letterpaper]{article} % DO NOT CHANGE THIS
\usepackage{aaai21}  % DO NOT CHANGE THIS
\usepackage{times}  % DO NOT CHANGE THIS
\usepackage{helvet} % DO NOT CHANGE THIS
\usepackage{courier}  % DO NOT CHANGE THIS
\usepackage[hyphens]{url}  % DO NOT CHANGE THIS
\usepackage{graphicx} % DO NOT CHANGE THIS
\usepackage{natbib}  % DO NOT CHANGE THIS AND DO NOT ADD ANY OPTIONS TO IT
\usepackage{caption} % DO NOT CHANGE THIS AND DO NOT ADD ANY OPTIONS TO IT
\usepackage{xcolor}

\usepackage{makecell}

\usepackage{caption}
\usepackage{subcaption}
\usepackage{graphicx}

\newcommand{\change}[1]{#1}
\newcommand{\minorchange}[1]{#1}

\title{From an Authentication Question to a Public Social Event:\\Characterizing Birthday Sharing on Twitter}
\author{
    Dilara Kek\"{u}ll\"{u}o\u{g}lu,
    Walid Magdy,
    Kami Vaniea
    \\
}
\affiliations{
    School of Informatics\\
    The University of Edinburgh\\
    Edinburgh, UK \\
    
    d.kekulluoglu@ed.ac.uk,
\{wmagdy, kvaniea\}@inf.ed.ac.uk
}

\begin{document}

\maketitle

\begin{abstract}
Date of birth (DOB) has historically been considered as private information and safe to use for authentication, but recent years have seen a shift \change{towards wide public sharing}. In this work we characterize how modern social media users are approaching the sharing of birthday wishes publicly online.
Over 45 days, we collected over 2.8M tweets wishing happy birthday to 724K \change{Twitter} accounts. For 50K accounts, their age was \change{likely} mentioned revealing their DOB, \change{and} 10\% were protected accounts. 
Our findings show that the majority of both public and protected accounts seem to be accepting of their birthdays and DOB being revealed online by their friends even when they do not have it listed on their profiles.
We further complemented our findings through a survey to measure awareness of DOB disclosure issues and how people think about sharing different types of birthday-related information. 
Our analysis shows that giving birthday wishes to others online is considered a celebration and many users are quite comfortable with it. This view matches the trend also seen in security where the use of DOB in authentication \change{process} is no longer considered best practice. 
\end{abstract}

\section{Introduction}

Online Social Networks (OSN) thrive on getting users to share information about themselves. One of the ways humans build relationships is through sharing of personal information to build trust, therefore it is unsurprising that they use social networks to do so. OSNs encourage sharing personal information by adding prompts like ``Wish Pat a happy birthday today!'' or ``Tell your friends about your new shoe purchase''. Even LinkedIn, which is \change{an OSN} focused on professional networking, has prompts about birthdays. 

Historically, date of birth (DOB) was considered as private information; this is why it has been used widely in authentication. Even today, some organizations such as phone companies and banks still use DOB as one of several authentication questions when users phone in~\change{\cite{kaur2020intelligent,lee2020empirical,twiliocall}}. The treatment of DOB as private data can also be seen in the European GDPR regulation where DOB is legally considered to be personal data~\cite{gdpr2} and is also regularly reported in data breach reports to the public as important personal information that may or may not have been lost during the breach~\cite{jackson2019investigation}.
Still, the use of DOB in authentication these days is significantly lower compared to a couple of decades ago before the spread of social media.  

Early research by Rabkin~\cite{rabkin2008personal} surveyed the password recovery mechanisms of 20 banks with the aim of showing how vulnerable they were. They found that DOB is used in the process of password recovery of some of these banks, and highlighted that this information can be inferred using public data found in OSNs, which opens the accounts to automatic attacks. This early study was the first to highlight that the trend of how DOB is seen is changing in the era of social media\change{, shifting from a fact that is shared only with close friends and family to a fact that is publicly shared with complete strangers online.}

In this paper, we characterize the disclosure of birthdays/DOBs on the Twitter platform and reactions of the users to such celebrations. We explore the tension between birthdays being open celebrations and birth dates as \change{private} information and investigate behavior by measuring the disclosure of birthday wishes on Twitter and users' reactions to them; and attitudes through a user survey asking Twitter users about their thoughts on the topic.

Our main research question is: \emph{Do social media users see their birthdays/DOB as private information anymore?}
More precisely we investigate the following sub-research questions:
\begin{description}
  
    \item[RQ1] How many birthday wishes on Twitter are posted publicly? What percentage of those indicate the \change{mentioned person's} exact date of birth?
    \item[RQ2] Do protected accounts, \change{who are theoretically more privacy concerned,} see less disclosure of birth days/dates than public \change{accounts}?
    \item[RQ3] How do Twitter users react to these public wishes? \change{Is there a difference in the reactions of public and protected accounts?}
    \item[RQ4] How aware/comfortable are Twitter users \change{with} having their birth day/date disclosed online? 
\end{description}

To answer our research questions, we \change{collected} over 18 million tweets/retweets mentioning ``happy birthday'' over 45 days. Of those, 2.8 million tweets directly mention one non-verified user account. 
The number of tweets shows just how many birthdays are being disclosed over a single OSN\footnote{\minorchange{For context, around 0.85\% of English tweets on Twitter contain the term ``birthday''. Based on a two week collection of Twitter's ``Sampled Stream'' which is a 1\% random sample of all tweets on Twitter.}}. 
Interestingly, we found that over 66K of these tweets \change{likely} disclosed the age of almost 50K unique users (e.g. ``\textit{Happy 16th birthday @user}''), which makes easy to directly infer the exact date of birth (DOB) of the user.
Some of the mentioned accounts were ``\textit{protected}'', where users have explicitly indicated that their tweets should be kept private and only visible to an approved list of people; yet these users still had their birthday, and sometimes DOB, publicly disclosed by their \change{followers}. 
While public account holders' ages are tweeted more often than the age of the users with protected accounts, still over 5K protected account holders' DOB were \change{likely} exposed within the 45 days of our collection period.

Finally, our user survey measured Twitter users' opinion/awareness on birth day/date exposures through celebration on the platform. 48\% of the participants were comfortable with others tweeting publicly about their birthdays including their ages.

Our findings indicate that indeed Twitter users are publicly expressing birthday wishes, sometimes also exposing the full DOB, even for protected accounts. The majority of the users are reacting positively to having their birthday and DOB disclosed publicly.
These findings show that the view of social media platform designers is the closest to the reality; \change{a large number} of users do not think that birthday and DOB are sensitive information anymore. This finding should be taken into account by the organizations that still use this piece of information in their authentication process.

%%%%%%%%%%%%%%%%%%%%%%%%%%%%%%%%%%%%%%%%%%%%%%
\section{Background}
\label{ref:background}

People use OSNs to share their experiences, interact with each other, as well as, to gain social capital~\cite{ellison2011negotiating}. Some people may use OSNs in professional contexts~\cite{mahrt2014twitter} and to build reputation~\cite{syn2015social}, while others may use them to seek social support from others~\cite{yin2016prayfordad}. Users maintain weak ties by interaction on social media~\cite{vitak2014facebook} and celebrating birthdays is one of the popular ways to do that~\cite{viswanath2009evolution}. However, these interactions can leak information if they are public. Individuals' privacy is connected to their networks in OSNs~\cite{boyd2012networked,amon2020influencing}, which can lead to unintentional disclosures~\cite{kekulluoglu2020analysing}. Most of the time, users cannot directly control these disclosures, and they might not even realize the reach of it~\cite{bernstein2013quantifying}. While there is research on utilizing these privacy leaks to get information on individual users~\cite{jurgens2017writer}, to our knowledge, there is no research focused on the characterising birthday and DOB disclosures by networks or the users' privacy concerns regarding this situation.

\subsection{Awareness of data sharing}

OSN users want to both share information and control its reach. However, given that users underestimate audience size~\cite{bernstein2013quantifying}, do not fully understand the visibility to  third parties~\cite{king2011privacy}, and have difficulty understanding that information shared online can result in other types of information being inferred~\cite{acquisti2006imagined}, an argument could be made that truly controlling information flow is quite challenging for an OSN user.

OSN users' privacy awareness has increased from the early days of Facebook~\cite{tsay2018social}, particularly in regards to their understanding of the visibility of their data to the public and the visibility to their connections. However, even with these improvements, people still struggle to understand how broad the reach of their posts are. Bernstein et al.~\cite{bernstein2013quantifying} looked at the true audience reach of 220,000 Facebook users as well as surveyed users about their perceived audience. They found that the imagined size of the audience was only 27\% of the true size.

\subsection{Sharing other people's data}

	The term ``networked privacy'' is pitched by
	boyd~\cite{boyd2012networked} to reflect on the collective aspects
	of privacy in social media. Even when individuals protect their private information, their networked relations may disclose them. An example would be a photo which is uploaded by a user onto an OSN and then tagged with the people in it. Each member of this collective (photo taker, photo subjects, event space owner) has some privacy stake in the photo and therefore its management is a collective issue. However, most OSNs give the right to manage the privacy settings of a content only to the uploader. While this enables users to control their self-disclosures, they have no say over what others share about them. Trusting privacy protection to the users' network might not be sufficient~\cite{pu2017valuating} and priming the network might even backfire, leading them to share more~\cite{amon2020influencing}. Some research has looked into how to handle this type of situations automatically on behalf of the user~\cite{kekulluoglu2018preserving,kokciyan2017argumentation}, but it is still in the proof-of-concept stage.

	Unlike offline interactions, there is a certain permanence to online posts and interactions. Hence, users might want to edit or delete some of their posts~\cite{yilmaz-2021-editingthepast}. However, some parts of the interactions can stay in the platform and leak information~\cite{kekulluoglu2020analysing}. For example, Twitter keeps the replies to a protected/deleted tweet visible in the platform.
	
	Some social media users choose to pause or stop using their accounts for various reasons including being ``in tune with'' themselves~\cite{baumer2018departing}, productivity~\cite{grandhi2019stay}, religious practices~\cite{schoenebeck2014giving}, as well as privacy protection~\cite{grandhi2019stay,lampe2013users}. However, users' friends can still share posts that lead to privacy violations~\cite{amon2020influencing,lampe2013users}. Even without sharing any posts, networks of non-users could still disclose enough information to create shadow profiles~\cite{garcia2018collective}.

 \subsection{Finding and inferring personal data}
Several studies have looked at the types of private information shared on OSNs as well as how to use that data to infer information which has not been shared. Mao et al.~\cite{mao2011loose} looked at the information shared deliberately on Twitter; they find that events such as vacation, illness, and drinking are shared. One type of information can also be used to infer more information. For example, burglars can use the above types of tweets to know that a user's house is vacant (e.g. \url{PleaseRobMe.com}). Insurance companies could also increase their premiums according to the severity of illnesses shared in the tweets~\cite{insuranceloss}.
Jain et al.~\cite{jain2013call} studied phone numbers posted publicly on Twitter and Facebook in India. They found that most of the phone numbers were intentionally posted by their owners. However, they were also able to use the phone numbers to find the name of the owner, voter ID, family details, age, home address, and father's name. By adding the numbers to WhatsApp they were able to get further information such as their US numbers, relationship status, and so on.

A user's connections on OSNs can also be used to learn quite a bit about the user, even if that user has ``locked down'' their account using settings. Jurgens et al.~\cite{jurgens2017writer} showed that an analysis of tweets mentioning a user is enough to determine their gender, age, religion, diet and personality traits. Kekulluoglu et al.~\cite{kekulluoglu2020analysing} found that life events such as marriage, graduation, surgery recovery of a user can be inferred by only looking at the replies sent to them. Analysis of OSN friend networks has shown that knowing information about a user's friends is sufficient to accurately infer attributes such as age, gender, location, political orientation, and sexual orientation~\cite{al2012homophily,zheleva2009join,jernigan2009gaydar,jurgens2013s,aldayel2019your}. Magdy et al.~\cite{magdy2017fake} inferred users' gender and age with 92\% accuracy from their network interaction and comments, which allowed them to spot ``fake'' accounts that might be used for catfishing on adult social networks.
Similarly, Garcia et al.~\cite{garcia2018collective} found that using networks of people on Twitter, allows detecting the physical location of a user with median error of 68.7km, and identify the city the user lives in with 32\% accuracy.

\change{\subsection{Birth dates in the authentication process}}

\change{Best practices advise against using knowledge-based questions in the authentication process~\cite{NIST} which includes asking for the birth dates as security questions. Usage of DOB should especially be avoided since it is considered easily discoverable information~\cite{canadaguide}, especially with the spread of social media~\cite{rabkin2008personal,irani2011modeling}. Even with these warnings, some organizations such as banks~\cite{kaur2020intelligent,murdoch2010verified,smyth2010forgotten}, wireless carriers~\cite{lee2020empirical}, and email service providers~\cite{li2017security,al2018web} still use DOB while authenticating users. 

Against the best security practices~\cite{euguide}, birth dates are also commonly used by people while constructing passwords~\cite{brown2004generating,bonneau2012birthday, wang2019birthday}. People also use DOB in their PINs which make them easier to be predicted. According to Bonneau et.al.~\cite{bonneau2012birthday}, lost or stolen wallets will lead thieves to correctly guess PINs up to 8.9\% of the time and the primary reason for that is the identification cards with DOB found in the wallets.}

%%%%%%%%%%%%%%%%%%%%%%%%%%%%%%%%%%%%%%
\section{Data Collection and Analysis Methodology}

We collected tweets containing the words ``happy'' and ``birthday'' and then analyzed them in regards to the amount of disclosure, type of account (public, protected), age disclosure, and engagement by the mentioned person.
In the following, we describe our data collection and annotation methodology that enables our initial quantitative analysis.

\subsection{Collecting tweets}

We used the Twitter streaming API~\cite{twitterapi} to collect \minorchange{public }tweets in real time. We filtered for English language tweets that contained both the words ``happy'' and ``birthday'', resulting in only tweets containing both those words, but not necessarily in consecutive order. 

We \change{collected} tweets for 45 days between January and March 2019 resulting in nearly 18 million tweets and retweets.  We filtered out the 11 million retweets as we are only interested in the initial birthday mention. In addition, we filtered out 2.3 million tweets that had no mentioned account along with 630K tweets mentioning multiple users where it was unclear whose birthday was disclosed. \change{For some accounts, Twitter will \emph{verify} the identity of the account holder and add a blue tick beside their user name. These tend to be owned by public figures rather than average users. Hence, w}e also removed 1 million tweets that mentioned verified accounts as well as 3K tweets where the user mentioned themselves. 
After this cleaning process, we ended up with a set of around 2.8 million tweets that use the words ``happy'' and ``birthday'', as well as mention only one non-verified account. We refer to this dataset as ``\emph{BD}'' tweets dataset.

Two days after the last tweet was collected, we batch processed all \emph{BD} tweets by: 1) identifying any mentioned accounts, 2) checking if the mentioned accounts are public or protected.
We excluded 44K tweets where the mentioned account could not be reached (e.g. deleted or suspended) at the time of processing. We then labeled each tweet in the collection with two labels in terms of:
\begin{enumerate}
    \item account status of the mentioned accounts: either mentioning a public account (\emph{mPublic}) or mentioning a protected one (\emph{mProtected}).
    
    \item tweet conversation type: 
    either a \emph{reply} or, \emph{directed to a user}. A \emph{reply} is in response to an existing parent tweet, such as when a user tweets about their own birthday and a follower replies. A tweet \emph{directed to a user} is a new tweet without a parent, mentioning the user (e.g. ``@username Happy 21st birthday''). These are likely to be wishes by friends of the mentioned user who already know their birthday.
    
\end{enumerate}

Its worth mentioning that \emph{protected} account tweets are visible to their approved followers only and cannot be retweeted or quoted by other users. However, if a \emph{public} account replies to a \emph{protected} account's tweet, the reply can be seen publicly. This is also the case for any tweet mentioning a protected account. Protected accounts can also be mentioned by non-followers.

\begin{table*}[t]
	\centering
	\footnotesize
	
	\begin{tabular}{ c c r r r } 
 
&\thead{\# tweets \\ (unique mentions)} & \thead{\% reply \\ to a user} & \thead{\% directed \\to a user} & \thead{\% with\\ two digits}\\
%[0.5ex] 
 \hline 
 \textbf{mProtected} & 202K (88K) & 30.5& 69.5   & 3.2 \\ %\hline
\textbf{mPublic} &2.6m (636K) & 42.7 &57.3  & 2.7 \\ %\hline
\hline
\textbf{Total} &2.8m (724K) & 41.8 & 58.2   & 2.8 \\
\end{tabular}
\caption{Overview of \emph{BD} dataset broken out by the type of account - mProtected and mPublic.}
\label{table:overview}
		\begin{tabular}{ c c c c c c c c } 
			%\hline
			%\hline
			&\thead{Tweets}  &\thead{\% Liked} &  \thead{\% Retweeted} & \thead{\% Replied}&  \thead{Avg time \\ to reply (h)} & \thead{\% Any \\ Reaction} & 
			\thead{\% not \\ accessible} \\
			%[0.5ex] 
			\hline %\hline
			\textbf{mProtected} & 5000 &  51.6 & 13.8 & - & - & 54.1 & 12.4 \\ %\hline
			\textbf{mPublic} & 5000 &  56.1 & 19.9 & 43.6 & 3.5& 66.6 & 8.5\\ %[2ex] 
			
			%\hline
		\end{tabular}
	%\end{adjustbox}
	\caption{Responses by the mentioned accounts. mProtected estimates were computed using the count of hidden interactions.} 
	\label{table:responses}
	\begin{tabular}{ c c c c c c} 
			&  \textbf{Accounts} & \textbf{No info} & \textbf{BD} &\textbf{BY} &\textbf{DOB}  \\
			\hline %\hline
			%\\[-1.5ex] 
			\textbf{Protected}  & 4159 & 3603 (86.6\%)  & 487 (11.7\%) & 30 (0.7\%) & 39(0.9\%) \\ %\hline
			\textbf{Public}  & 4363  & 3474 (79.6\%) & 763 (17.5\%) & 43 (1\%) & 84 (1.9\%) \\ %[1ex] 
		\end{tabular}
		\caption{Birthday information sharing patterns by the type of account - protected and public.}
		\label{table:profile}
\end{table*}

\subsection{Gathering reactions on BD tweets}
We also measure the reaction of the mentioned user accounts to birthday tweets. For measuring the reactions, we collected the engagement of the mentioned account with these tweets either by replying to or liking the tweet.

Inspecting all these tweets individually to check interaction with them was impractical, due to Twitter API limitations. Thus, we randomly sampled a set of 10,000 tweets (5,000 mPublic, 5,000 mProtected) from the BD dataset. To avoid bias, we took samples equally from each day. We refer to this sample of our dataset as \emph{BD-react}.

Twenty days after the last tweet in our main data set was collected, we measured the amount of engagement tweets in \emph{BD-react} had experienced. The average time to reply to a tweet in our set was 3.5 hours, so we are fairly confident that the majority of engagement will have happened within our 20+ day time period. 
For protected accounts, it is not possible to see if the account has interacted with a tweet via API. However, how many protected accounts have retweeted or liked a tweet is visible via Twitter's user interface (UI). Thus, for mProtected tweets, we scraped if they have been liked or retweeted by a protected account. \change{Note that we can only understand whether \emph{a} protected account interacted with the tweet but we cannot get the usernames of those users to check whether the interaction was by the mentioned protected account.}
To determine if any of the mentioned accounts had replied to the tweet, we collected the tweets of the mentioned account and searched for replies to our recorded tweet. Doing so was necessary because the Twitter API does not have a method to collect replies to a particular tweet. After this process, each tweet in our \emph{BD-react} was labeled as being liked, retweeted, and/or replied to (in the case of mPublic) by the mentioned user.

\subsection{Gathering Birthdays on Profiles}

	While our main focus is on birthday disclosure by others, 
	users themselves might be self-disclosing the information publicly on their profile pages. In this case, the user may be fine with birthday exposure and others might feel encouraged to tweet about their publicly visible birthday. 
	To see whether users shared their birthday or date information in their profiles, we collected the public birthday information from each account. This information could be gathered for both public and protected accounts. We collected the self-disclosed birthday information for all accounts in our \emph{BD-react} collection. We have applied this process a few months after our initial collection, which led to losing access to some of the accounts due to deletion, deactivation, or suspension; resulting in getting the information of only 4364 public and 4159 protected unique accounts.

\subsection{Tweets disclosing user's age}
Some tweets explicitly mention the age of the person. An example from our BD dataset (username anonymized): ``Happy 40th Birthday to @username. Have a great day and night''. 
If the age is combined with the date the tweet was posted, it becomes trivially possible to reconstruct the full birth date. To understand the scope of this disclosure, we further analyzed the tweets to extract those containing a two-digit number between 10 and 99. We looked for all instances of two digits on their own or in combination with an ordinal indicator (i.e. ``st'', ``nd'', ``rd'', ``th''). We selected the 10-99 range, because numbers below 10 might mean something other than the age, and technically Twitter does not allow users younger than 13 years old. Similarly, few people live to over 99, so the number of errors in this numeric range is expected to be large compared to the number of true ages. The percentage of tweets that contain two digits with our criteria are shown in Table~\ref{table:overview}.

To verify if the tweets containing two-digit numbers are referring to the user's age, we manually labeled a random sample of of 4000 tweets (2000 mPublic, 2000 mProtected) from the tweets that had two-digit numbers. We took samples equally from each day as we did with \emph{BD-react}. We refer to this sample of tweets as \emph{BD-age}.

For the annotation, we used the online crowdsourcing platform Appen\footnote{\change{Appen is f}ormerly known as Figure-Eight and Crowdflower. \url{https://www.appen.com/}.} participants were asked: ``Can we tell that this person: @username has their age disclosed in the tweet?'' where @username was replaced with the mentioned person's account from the actual tweet. 
Each tweet was judged by three trusted workers and we used majority voting to label tweets. A test-set of 64 pre-labeled tweets, that we manually annotated, was provided for quality control of the annotation. If a worker got more than 20\% of the pre-labeled tweets incorrect, their annotations were discarded. The final inter-annotator agreement rate was 93\%.

\change{\subsection{Ethical considerations} 
This work aims to measure the prevalence and practice of birthday celebrations on Twitter, a question that is most practically answered by sampling public tweets from Twitter itself. However, such sampling comes with some ethical conundrums since we are using data for research purposes that was initially shared presumably to connect with others. We are also aware that some users publicly mention protected accounts, potentially sharing information, like DOB, that the protected user would prefer to keep private. }

\change{To limit potential negative impacts of our work we take several steps. We only collect publicly available tweets. In the paper we report on aggregate information and refrain from singling out individuals in quotes, links, or anything else identifying. When using quotes, we carefully select those that are generic and represent common tweet content (i.e. ``Happy sweet 16th birthday!''). We also do not collect our survey participants' Twitter data or link it to their answers. We received ethical approval from our institution for this work, including the Twitter data collection and the following user survey.}

%%%%%%%%%%%%%%%%%%%%%%%%%%%%%%%%%%%%%%
\section{The Share of Birthday Wishes on Twitter}

\begin{table*}
\centering
\footnotesize
 \begin{tabular}{ c c c c c } 
& \thead{Accounts} & \thead{Received \\ two-digit tweet} & \thead{Age likely \\ known } & \thead{\% age likely \\ known }  \\
%[0.5ex] 
 \hline %\hline
 %\\[-1.5ex] 
 \textbf{Protected}  & 88K & 5.5K & 5K & 5.7 \\ %\hline
\textbf{Public}  & 636K  & 51K& 44K & 7 \\ %[1ex]
%\hline
%\hline
\end{tabular}
%\end{adjustbox}
\caption{Estimations of the number of accounts where the full birth date and year can be determined. Estimations are based on a combination of the number of tweets containing a two-digit number and the observed rates of age prediction from the Appen annotations.}
\label{table:reflection}
\end{table*}

\begin{figure*}
   \centering
    \includegraphics[width=0.7\linewidth]{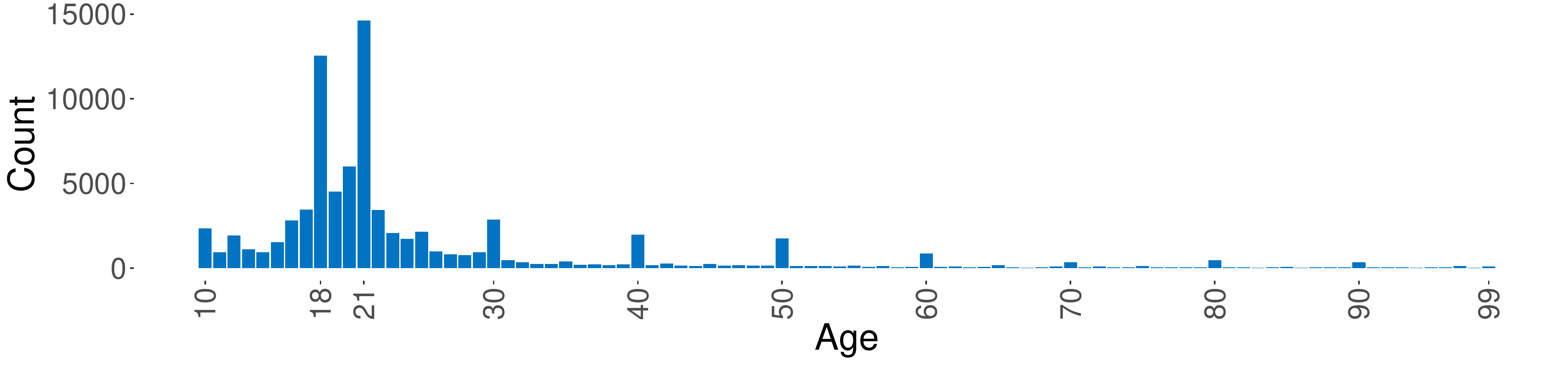}
    \caption{Distribution of the two-digit numbers in \emph{BD}. Notable spikes at key ages: 18, 21, and multiples of 10.}~\label{fig:ageDist}
\end{figure*}

\subsection{BD dataset statistics}
Table~\ref{table:overview} shows the overview of our \emph{BD} dataset, which contains 2.8m original tweets, broken out by the type of the mentioned account - mProtected, mPublic. 
These 2.8m tweets were directed to 724K unique accounts. The majority (56\%) of these accounts received only one birthday celebration tweet. 99\% of them received 30 or less birthday wishes. Protected accounts tended to get less birthday wishes on average with 99\% of them receiving 15 or less tweets. One public user in our collection received 3934 birthday wish tweets, while the most popular protected user received 394 birthday wishes.

We also check the percentage of tweets that were replies to other tweets or written to a user directly. 
We see that tweets were more likely to be directed to a user without replying to an existing tweet.
Only 42.7\% of the tweets were a reply to an existing tweet, whereas 57.3\% of them were directed to a user without the user tweeting about their birthday. This difference is even higher for mProtected, where 69.5\% were tweets directed to them and only 30.5\% were replies (Table~\ref{table:overview}).

Finally, looking at the tweets containing two-digit numbers (including ordinal indicators) mProtected tweets had higher percentage  (3.2\%) than the mPublic tweets (2.7\%). 
These numbers can be an indication of the mentioned person age, which can lead to easily inferring the person's exact DOB. We provide further analysis for the meaning of these numbers later.

\subsection{Twitter users' reactions to birthday wishes}
\label{sec:reactions}

We carried out another analysis on \emph{BD-react} in which the tweets were processed in more depth to analyze the reaction of the mentioned users to the BD tweets mentioning them (Table~\ref{table:responses}). 
At the time of processing, 10.5\% of these were no longer accessible due to various reasons such as the deletion of the tweet, the author protecting their account, and so on.
mPublic tweets received high interaction from the mentioned accounts. We observed that 56.1\% of the mPublic tweets had likes from the mentioned account. 
For mProtected tweets, we only know that a protected account interacted  with the tweet, not which account.  However, hidden interactions can give us an idea.
51.6\% of mProtected tweets had hidden likes while 13.8\% of them had hidden retweets.
66.6\% of the public mentioned accounts interacted with the tweets mentioning them while 54.1\% of the mProtected tweets had hidden interactions. This result shows that people frequently interact with the tweets that wish them a happy birthday in a positive way such as liking and retweeting, regardless of those people's accounts being public or protected. In addition, the large number of interactions can indicate that the tweets are seen by other people who might not be necessarily following the birthday person.

	\subsection{Sharing Birthdays on Profile}

	By checking the birthday information on the profiles of the 8522 reachable accounts, we found 7077 (83\%)  shared no birthday information, 1250 (14.7\%) shared the birthday (BD), 73 (0.9\%) shared only the birth year (BY), and  123 (1.4\%) shared their full DOB.
	Public accounts were more likely to share information on their birthday (890, 20.4\%) than protected accounts (556, 13.4\%).
	In total only 196 (2.3\%) of the users disclosed their birth year. 
	We report the birthday sharing behavior on profiles broken out by the account type in Table~\ref{table:profile}.
	Users who shared their birthday information reacted similarly to the birthday tweets with those who did not. 
	This was also the case for protected users.

\subsection{DOB leakage on Twitter}

Regarding the \emph{BD-age} tweets, 82.9\% of the tweets with two-digit numbers refer to the mentioned person's age, according to Appen annotators. The percentage is slightly higher for the mentioned protected accounts (84.4\%) than public ones (81.3\%). We noticed that this percentage becomes higher (95.3\%) if the two-digit number is followed by ordinal indicator (st, nd, rd, th). Using these rates, we extrapolated to the whole data set, taking into account the total number of tweets containing two-digit numbers, the results are shown in Table~\ref{table:reflection}.

We look at the unique accounts mentioned in the \emph{BD} dataset to understand the potential DOB disclosure for birthday people. There were 56K (8\%) accounts in total that received at least one birthday tweet that contained a two-digit number, of those 33K received at least one tweet accompanied by an ordinal indicator. 51K of them were public accounts, while 5.5K of them were protected accounts. Based on the results of the annotation, we can estimate that the actual age of the person is exposed for over 49K accounts which when combined with the date of the tweet, likely exposed the full birth date and year. This is 6.8\% percent of the accounts that were mentioned in the tweets we collected.

The mean of the two-digit numbers we found is 25 with median 21. The most celebrated ages were 18 and 21, followed by ages at multiples of ten (Figure~\ref{fig:ageDist}). 
From our collection, we see that over 1K accounts receive birthday wishes that exposes their DOB every day, where 10\% of those are protected accounts. While these users are mostly young adults, there are also users who are teenagers and elderly. Public accounts got more age exposing tweets than the protected accounts which suggests that people treat accounts differently depending on their type.  Interestingly, accounts that shared no birthday info got more birthday messages with two-digits.

Combining these results with the reaction of those users on the tweets, it becomes necessary to understand how Twitter users see this phenomena and if they perceive the DOB as private information.

%%%%%%%%%%%%%%%%%%%%%%%%%%%%%%%%%%%%%%%
\section{Measuring Users' Opinions and Awareness}

We conducted a survey to better understand how Twitter users think about the public sharing of birthday wishes on Twitter (RQ4), as well as their understanding of tweet visibility settings.
We advertised the survey on Prolific Academic (PA)~\cite{prolific} as ``Wishing a Happy Birthday on Social Media''. 
The advertisement limited participants to Twitter users from the United States or United Kingdom to ensure similar culture and English label proficiency. 

We followed our University's ethics protocol in the design and running of the survey. Participants were compensated £0.5 (£8.34 per hour).

\subsection{Survey Instrument}
The survey started with informed consent followed by a screening question about if they had a Twitter account and if they used it more or less than once a month. Those without a Twitter account were screened out.
We then asked if their primary Twitter account, was public, protected, or sometimes protected where they change the settings, followed by if they associated their Twitter account with their ``real identity'', and if they 
had their birthday publicly visible on \emph{any} social media account.

To gauge understanding of Twitter setting impacts, we asked what would happen in two scenarios where public and protected accounts interact. 
We also asked if they can tell that a poster's account is public or protected when replying to a tweet, and if they look to see if the account is protected when engaging with tweets (reply, mention, retweet). 

To understand their comfort with public birthday and date disclosure we asked them how comfortable they would be with friends and family publicly tweeting about their birthday with and without age information. We also asked how they might engage with such a tweet (like, retweet, reply, direct message (DM), ask to remove).
We then asked them a similar question around the participant tweeting about a friend or family member's birthday with and without age. 
\change{Finally, to gauge participants' understanding of the positive and negatives of public birthday wishing we asked them two free-text questions: \emph{``Give at least one example of a good thing that could happen if someone knew your birthday and age.''} and the same question with \emph{``bad thing''}.} 
The survey ended with an optional comment box. 

As PA provides common participant demographics to researchers, we did not directly ask for any demographics.

\begin{figure}
\captionsetup{justification=raggedright}  
\centering
  \includegraphics[width=0.85\linewidth]{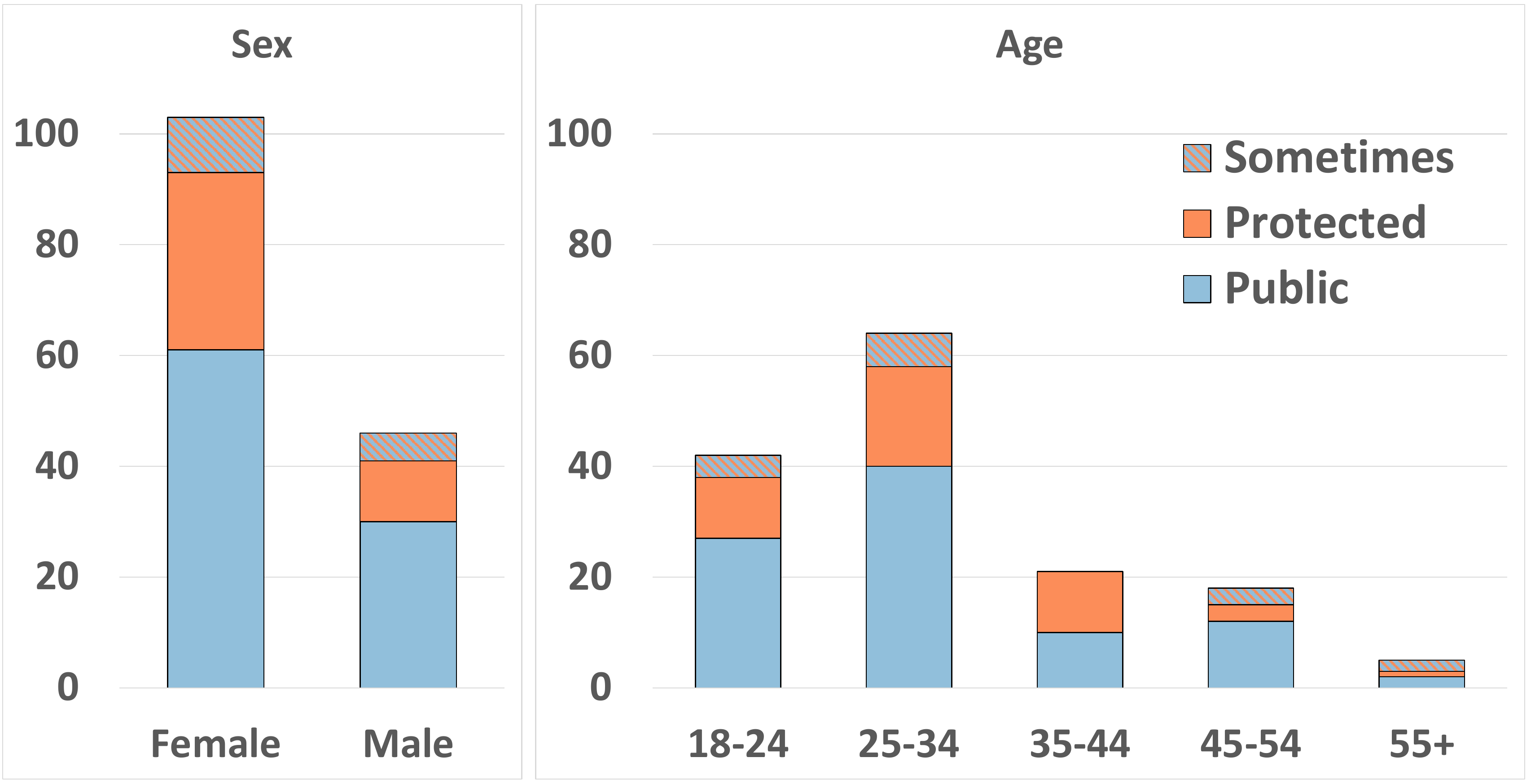}
  \vspace{-0.3cm}
   \caption{\change{Sex and age distributions of survey participants by account type. ``Sometimes'' refers to accounts switching between public and protected.}}
   \label{fig:ageAccount}
\end{figure}

\subsection{Survey results}

\paragraph{Participant demographics}
The survey had 151 participants, 118 from the United Kingdom (UK) and 33 from the United States (US). Their average age was 31.2 years ($\sigma=10.6$) with a median of 28.5. 
Respondents were primarily female (n=103, 68.2\%) vs.\ male (46, 30.5\%) with 2 preferring to not respond.

\change{For context, Twitter users from US have median age of 40 and half of them are female~\cite{wojcik2019sizing}. Only 44\% of the UK Twitter users are female~\cite{twitterukgender} with more than half of the users older than 35~\cite{twitterukage}. Our participants are generally younger than the general Twitter population and have a higher percentage of female representation.}

\paragraph{Account types} Most participants had public accounts (92, 61\%) with the rest having protected accounts (44, 29\%) or switching between public and protected (15, 10\%). Figure~\ref{fig:ageAccount} illustrates sex and age distributions of the survey participants broken out by their account types.
84 (55.6\%) participants associated their Twitter accounts with their real identity. Of those, 27 (32\%) had a protected account or switched between public and protected (9, 11\%). In other words, roughly half of the people whose accounts were linked with a real identity were also protected. 
Publicly listing a birthday on at least one social media account was common, with 96 (63.6\%) publicly listing a birthday, 46
(30.5\%) not listing, and 9 (6\%) not sure.
65\% of the participants with public accounts shared their birthday on at least one social media account, whereas 57\% of the protected accounts shared it. 75\% of the users who switch between public and protected shared their birthday on social media.

\change{13\% of the adult US Twitter users have protected accounts~\cite{wojcik2019sizing}, which is much lower than the share of protected accounts in our survey participants.}

\paragraph{Visibility of protected accounts}
Participants were asked about two scenarios: imagine that \textit{``Alice (Public) retweeted one of Bob's (Protected) tweets using the Twitter website?''} and \textit{``Alice (Public) tweeted at Bob (Protected) using his handle (@bob) in the tweet?''} We then asked them if Twitter would: allow Alice, warn Alice, allow Alice but restrict visibility to Bob's friends, allow Alice with no restriction, or they didn't know. 
For retweets, the Twitter website does not allow public accounts to retweet protected accounts. Only 45 (29.8\%) of participants gave this answer. The most common answer was that Alice could retweet but only Bob's followers could see it (70, 46.4\%), which is incorrect. 

For tweeting at protected accounts, the Twitter website allows public accounts to mention protected ones with no visibility restrictions. 
Participants were split on this question with 58 (38.4\%) thinking that the tweet would be visible (correct), and 58 (38.4\%) thinking that it would be visible only to Bob's followers (incorrect).
The confusion over how Twitter protected accounts work is further highlighted by comments from participants about why they switch their accounts between protected and public.
``\textit{If I post things I don't mind everyone seeing I changed it to public but if [I] post photo that I only want my followers to see I make it private.}''
The comment highlights a potential misconception that the protections are per-post instead of per-account.

When an account is protected, a padlock appears beside the username. However, when asked, only 33.8\% of the participants agreed that they can easily see whether the Twitter account
they are replying to is protected or not. Even less 
check the type of account they are replying to (24.5\%).

\paragraph{Birthday tweet opinions}
Looking at participants' comfort with friends and family tweeting about their birthday, the majority of participants were comfortable with their birthday being tweeted (103, 68.2\%). Most gave the same answer for birthday and birthday with age (90, 59.6\%) , indicating that the addition of age had no impact on their comfort. A further 58 (38.4\%) indicated that they would be less comfortable with birthday tweets containing an age.

Looking at participants' likeliness of tweeting about a friend or family member's birthday, the majority of participants indicated they would be unlikely to do so (79, 52.3\%), vs likely to do so (53, 35.1\%). 
The majority of participants (80, 53\%) gave the same answer for both birthday and the birthday with age, again indicating that the addition of age had no impact on tweeting likelihood. The rest (69, 45.7\%) were less likely to post a birthday tweet containing age.

Regarding their reactions to birthday wishes online, 90 (60\%) participant said they would reply with a thank you, or like the tweet (84, 56\%). Replying via direct message was less common (18, 12\%). And it was rare to ask the person to remove the tweet (7, 4.6\%) or retweet it (7, 4.6\%). 
Those with public and protected accounts indicated similar reactions to tweets. These findings are similar to Table~\ref{table:responses} where likes and replies were the main reactions to birthday wishes.

%%%%%%%%%%%%%%%%%%%%%%%%%%%%%%%%%%%%%%%%%%%%%%%%%%%%%%%%555
\paragraph{\change{Good \& bad impacts of sharing}}

\change{We asked our participants to list some of the  good and bad things that could happen if someone knew their birthday and age. Two researchers read through all the free-text answers and jointly grouped them into themes, discussing throughout to reach agreement. The most common good things mentioned were getting birthday wishes (45\%) and gifts (31\%). For the bad things, the most prominent worry was identity theft (37\%), and the next most mentioned  was being harassed, ridiculed or harmed (19\%). 8.6\% of the participants said nothing bad would happen.}

\change{The responses of participants on our survey showed that users are aware of the role birthday wishing can have in connecting them with others as well as some of the dangers of birthday and DOB disclosure can cause. Nevertheless, nearly half of them are still comfortable with their friends sharing their age along with birthday tweets online.}

%%%%%%%%%%%%%%%%%%%%%%%%%%%%%%%%%%%%%%
\section{Discussion}

\subsection{Summary of findings}

Our first research question concerns the number of birthday wishes posted on Twitter and how many include information revealing the DOB of the mentioned user. We collected 2.8 million happy-birthday tweets mentioning 724K unique users over 45 days, which shows that a large number of happy-birthday tweets can be easily linked to a specific account.
Further, we found that around 7\% of those accounts received at least one tweet that disclosed their age, allowing for easy combination of posting date and age to compute DOB.
Considering only around 2\% of the users \change{in our dataset} share their birth years on their profiles, DOB is being actively exposed for users who did not proactively share it.

Our second research question concerns how birthday wishes differ towards protected and public accounts.
88\% of all the birthday wishes we collected were towards public accounts. These accounts also received more birthday wishes per account on average than protected ones. We found that public accounts are also slightly more likely to receive birthday tweets that reveal their DOB than protected accounts, 8\% vs. 6.2\% respectively (Table~\ref{table:reflection}). The result indicates that having a protected account does not prevent online disclosure of birthday or date information. 

Finally, we measured users' reaction behavior (tweet interaction - RQ3) as well as their attitudes (survey - RQ4). 
We found that 66.6\% of tweets mentioning public accounts were reacted to by the mentioned account and 54.1\% of tweets mentioning a protected account were likely reacted to, though we have limited visibility of protected accounts. \change{Both Twitter and survey data show that liking and replying are popular ways to react to birthday tweets. 56\% of the mPublic tweets in \emph{BD-react} received a like from the mentioned person. Similarly, 56\% of the survey participants said they would like a birthday tweet they receive on Twitter. 60\% of the survey participants would reply to the birthday tweet, while 44\% of the mPublic tweets in \emph{BD-react} received replies from the mentioned user. Our Twitter data collection (20\%) and user survey (5\%) differ on the cases of retweets. 5\% of our survey participants selected that they would ask the person to remove the tweet if they received a birthday celebration over Twitter. While it is not possible to measure this reaction from the Twitter data directly, we recorded 428 (9\%) cases where the tweet or the user was deleted while collecting the replies to the tweets in \emph{BD-react}. However, there is no way to differentiate from the API response whether it was the tweet that was deleted or the user's account. There is also no way to make sure that the birthday person requested the tweet deletion in any case.}

These findings suggest that users are aware of happy birthday tweets and react to them positively. 
Our survey verifying our tweet analysis findings show that Twitter users are comfortable with public celebrations of birthdays, both with and without explicit mention of their ages. This result is evident both in the scale of current birthday wishing on Twitter as well as the attitudes of survey respondents. However, most of the respondents were less likely to publicly celebrate birthdays of their friends and family with tweets containing their ages.

Our survey also showed that Twitter users \change{might not be} fully understand\change{ing} who can see their tweets when they mention protected accounts and sometimes not aware of account types of users they interact with. We further discuss these findings in the implications section below.

\subsection{Limitations}
Our research analysis was limited to tweets on Twitter which explicitly
mentioned both the words ``happy'' and ''birthday''. While the term 
``happy birthday'' is culturally quite common in English speaking 
countries, there are many other ways to express the sentiment. 
Additionally, we did not look for common misspellings or 
abbreviations such as  ``hbd'', ``happy bday''. 
Therefore, the numbers presented in this work should be seen as
a lower bound, or  what an opportunistic data gatherer 
might be able to locate easily.

Another limitation is that we only look at the two-digit
numbers that have a leading space and no trailing alphanumeric characters other than the ordinal indicators. Because
of this, we are missing some age exposing tweets. 
Some tweets may also have
the age spelled instead of writing with numbers like ``twenty first'',
or ``sixteen''. 
We ran a test on \emph{BD} data set (Table~\ref{table:overview}),
to understand if people spelled out ages. We looked at a commonly celebrated birthday ``twenty one'' or ``twenty first''
and compared the occurrences with the numeric ``21'' and ``21st''. We  
found that $<0.01$\% of our \emph{BD} tweets spell out 21, likely due to the character limit pressures imposed by Twitter. Hence, we expect the effects of excluding spelled out numbers to be minimal.

%%%%%%%%%%%%%%%%%%%%%%%%%%%%%%%%%%%%%%
\subsection{Implications}

\paragraph{Popularity of birthday wishing}
Twitter is clearly considered a suitable platform to wish someone else a ``happy birthday'', which is further reinforced by verified account reactions as well as Twitter itself~\cite{twitter-birthday}. In our initial data collection we observed about a million tweets wishing a verified account a happy birthday. While we excluded these from analysis, the size of the dataset speaks to the public reinforcement that birthday wishing is a normal public Twitter activity. Twitter itself also encourages birthday celebrations by displaying a balloon animation on their birthday when others visit it. Other OSNs, such as Facebook, also actively encourage people to wish others a ``happy birthday'', which can impact the number of birthday wishes~\cite{birthday-devalued}. 

Birthday wishing was also common among non-verified accounts, accounting for nearly three quarters of the wishes. 
As an OSN, one of Twitter's roles is in helping people maintain weak and strong ties through sharing information~\cite{vitak2014facebook} which can have benefits on their mental health and sense of belonging~\cite{dym2018vulnerable}. These ties are also useful at helping people gain access to prospects such as jobs and opportunities~\cite{hoyle2017viewing}, so maintaining them has value. Viswanath et al.~\cite{viswanath2009evolution} found that users who do not interact on social media frequently, mostly only exchange birthday messages. Hence, birthday wishes can support and encourage users to maintain their social ties.

\paragraph{Disclosure control}
Even if a user is inclined towards not sharing their DOB on Twitter, they have limited control over their network. Most OSNs only provide control to the poster, not the data subject. Hence, \emph{networked privacy} means that the control over who can see what data is not solely in the hands of the person whose data it is. While some OSNs allow subjects to remove their tags from a post, they do not allow complete removal of the post. Consequently, if a user would prefer their birth date not be known, they would need to ask each poster to remove their post. Such a request may be socially challenging, especially if the majority of users feel that wishing someone a happy birthday is a good thing and nothing to be concerned about. Such a request might also risk being labelled as ``paranoid''~\cite{gaw06}.

Controlling information disclosure is also dependent on Twitter users and their networks' accurate understanding of tweet visibility. \change{Previous research on social media with granular privacy options show that users find it hard to comprehend and configure these settings~\cite{madejski2011failure}, while also underestimating the audience size~\cite{bernstein2013quantifying}.} Twitter has a relatively simplistic access-control approach for a modern OSN. An account is either public or protected. If public, anyone on the internet can see the posts, if protected, then only a selected set of users can see them. Yet even this simplistic model confuse\change{d our participants}. In our study we found that 38.4\% of participants thought that posts in reply to a protected account would also be protected. Findings suggest that users' mental models of Twitter protections \change{might be} inaccurate which may lead them to disclose information about protected accounts while honestly believing that the posts are not publicly visible. 

\paragraph{Authentication question vs. celebration}
There are surprisingly few questions that work well for authenticating identity. A good authentication question should: apply to nearly everyone, have a large number of possible answers, have equal distribution of answers, be easy to remember, and should not change. Based on those requirements, it is obvious why facts like birthdays were regularly used as a part of authentication. However, with birthdays celebrated publicly on OSNs and reaching more people than before, organizations justifiably shifted their use of DOB.

DOB is also used as one of the attack vectors in social engineering and re-identification methods~\cite{sweeney2000simple,krombholz2015advanced}. People often use their DOB when constructing passwords~\cite{brown2004generating}. Bonneau et.al.~\cite{bonneau2012birthday} showed that up to 8.9\% of the time, lost or stolen wallets will lead to thieves to correctly guess PINs. A primary reason is that DOB can be obtained using identification cards found in the wallet. However, as we can also see from our Twitter data and the following user survey, birthdays are not seen as secret by the general public. \change{Similar with our findings, Markos et al.~\cite{markos2017information} also found that DOB was considered as a low-privacy segment data along with e-mails by their participants compared to mother's maiden name, home addresses, and phone numbers. This attitude by public leads }to a tension between data privacy and the reality of cultural sharing. 
One of the main recommendations that we can learn from our study is that organizations, such as \change{banks~\cite{kaur2020intelligent,murdoch2010verified}, email service providers~\cite{li2017security,al2018web}, wireless carriers~\cite{lee2020empirical},} that still consider DOB sensitive information should withdraw from using it as a part of their authentication system. Users should also not incorporate their birthdays into constructed passwords.

\paragraph{Twitter design implications}

Birthday sharing is widespread on Twitter and users are comfortable with it. Encouraging this behavior might help users to feel valued and appreciated, as well as maintain friendships~\cite{vitak2014facebook}. In order to encourage birthday wishing, Twitter could notify followers of a user on their birthday if the user wants to receive such messages. Displaying a small indicator (e.g. balloon, cake) next to the username of a person having their birthday on their tweets may also act as a reminder for the birthday and encourage a message. Providing users personalized messages and collating the birthday tweets in one thread will help the birthday person to feel special and also allow easy access for replying to those messages. 

On the other hand, some people might not want their birthdays to be celebrated publicly. However, Twitter accounts cannot manage the tweets that mention/quote/retweet them. Some users solve this by asking other users explicitly not to interact. For example, some users state that they do not want other users to quote them in their profiles or usernames. Information about protected accounts can also be revealed through public replies to their tweets. Twitter recently introduced a feature to select user groups who could reply to specific tweets. While this is a positive step, these tweets are still publicly visible. Another recently added feature is the option to hide replies, however, it is easy to access these tweets with an extra click. They are also reachable by search. In addition, users might be socially uncomfortable to hide celebratory messages that leak information about them. Hence, users should be given control on the visibility of the tweets that mention them.

Another point is the confusion over the visibility of the posts when public and protected accounts interact. Our survey showed that users' mental models of tweet privacy \change{might be} quite different from the actual Twitter functionality. \change{Our participants} expect\change{ed} these tweets to be visible only to the followers of a protected account when a public follower interacts with the protected account. Hence, it is essential to disambiguate the interactions between them. This can be achieved by having an indication when interacting with protected accounts and let users know who can see the tweet if posted. The account types of the users mentioned in a tweet should be easy to check. As of now, when replying to a protected tweet, there is a padlock near the username indicating the status. However, when drafting a new tweet, the public or protected status is obscured.

\section{Conclusion}
In this study, we investigated the sharing of birthday wishes on Twitter, how they can reveal the date of birth of some users, and privacy concerns of the Twitter users regarding the DOB disclosure.
Our objective was to provide an in-depth analysis of how social media users see their DOB, as a private personal information, or as a happy event to be celebrated publicly. Our aim was to assist designers in the security and social media field to get a clear answer of how to treat this information about users when designing their systems.
We both conducted an analysis of 2.8 millions tweets sharing birthday wishes, and a survey of Twitter users to understand their opinions around public celebration of birthdays on the platform. We found that birthday celebrations are common over Twitter and over 1K tweets disclose the DOB of the mentioned account daily, where 10\% of those are protected. While the majority of those accounts do not share their birthday publicly, they still seem to be comfortable with others sharing birthday tweets publicly regardless of their account type, even when it discloses their DOB. We show that birthdays and DOB are not considered as sensitive information by users; they are celebrated publicly. Our findings should move any organization that is still using DOB as a part of their authentication process to phase its use out.

\bibliography{references}

\end{document}